# Sub-Ensemble Correlations as a Covariance Geometry

*Zuoxian Wang[1,2,4], Yuhao Zhang[1,2,4], Gaopu Hou[1,2,4], Zihua Liang[1,2,4], Gen Hu[1,2,4], Lu Liu[1,2,4], Yuan Sun[1,2,4], Feilong Xu[1,2,4], and Mao Ye[1,2,3,4],\**

1. Key Laboratory of Ultra-Weak Magnetic Field Measurement Technology, Ministry of Education, School of Instrumentation and Optoelectronics Engineering, Beihang University, Beijing 100191, China
2. Zhejiang Provincial Key Laboratory of Ultra-Weak Magnetic-Field Space and Applied Technology, Hangzhou Innovation Institute, Beihang University, Hangzhou 310051, China
3. Hangzhou Institute of Extremely-Weak Magnetic Field Major National Science and Technology Infrastructure, Hangzhou 310051, China
4. Hefei National Laboratory, Hefei 230088, China

*E-mail: maoye@buaa.edu.cn

**Keywords:** atomic sub-ensembles, correlation, covariance operator, spin fluctuations, stochastic field, spatial sampling overlap

**Abstract**

Conventional practice of spatially resolved detection in diffusion-coupled thermal atomic vapors implicitly treat localized responses as mutually independent. However, in this study, it is shown that observable correlations are governed by the intrinsic spatiotemporal covariance of a global spin-fluctuation field, such that spatial separation specifies only overlapping statistical projections rather than independent physical components. A unified field-theoretic description is established in which sub-ensembles are defined as measurement-induced statistical projections of a single stochastic field. Within this formulation, sub-ensemble correlations are determined by the covariance operator, inducing a natural geometry in which statistical independence corresponds to orthogonality of the measurement functionals. For collective spin fluctuations described by a diffusion–relaxation Ornstein–Uhlenbeck stochastic field, the covariance spectrum admits only a finite set of fluctuation modes in a bounded domain, imposing an intrinsic, field-level limit on the number of statistically distinguishable sub-ensembles. The loss of sub-ensemble independence is formalized through the notion of spatial sampling overlap, which quantifies the unavoidable statistical coupling arising from shared access to common low-order fluctuation modes. While multi-channel atomic magnetometry provides a concrete physical setting in which these constraints become explicit, the framework applies generically to diffusion-coupled stochastic fields.

## 1. Introduction

In diffusion-coupled thermal atomic vapors, collective spin dynamics is most naturally described as a stochastic field governed by thermal diffusion, relaxation, and microscopic noise [1,2]. Its evolution gives rise to intrinsic spatiotemporal correlations, so that any experimentally accessible signal probes a statistical projection of this correlated fluctuation field. Within spin-noise spectroscopy (SNS), previous work has primarily focused on how stochastic transport processes—such as diffusion, relaxation, and collisional exchange—shape the local temporal autocorrelation of spin fluctuations probed by a spatially confined measurement weighting [3–7]. In this context, it has been recognized that equilibrium spin fluctuations encode transport-induced spatiotemporal correlations [8–10], and that SNS can, in principle, access such correlations in both the temporal and spatial domains [11].

What has remained largely unexplored, however, is how correlations between spatially separated 'sub-ensembles' arise when multiple spatially resolved measurements simultaneously probe a common diffusion-coupled stochastic field. This is of direct practical relevance, as spatially resolved detection schemes are widely employed to extract multiple spin signals from distinct regions of a single vapor cell, particularly in multi-channel atomic magnetometry and related sensing platforms [12–21]. Implicit in such approaches is the assumption that signals obtained from geometrically separated probes are independent. In a diffusion-coupled medium, however, this assumption lacks a rigorous theoretical foundation. Continuous atomic motion driven by thermal diffusion prevents the long-term localization of spin-carrying atoms, undermining any purely geometric association between spatial separation and signal independence. This raises a question of principle that has yet to be systematically addressed: how should the correlations between spatially resolved spin signals be rigorously defined and quantified?

Addressing this question requires a conceptual reassessment of what constitutes a 'sub-ensemble' in a diffusive system. In the absence of dynamically persistent atomic subsets, spatially localized responses cannot be identified with independent physical subsystems. Rather, they must be understood as distinct statistical projections of a single global spin-fluctuation field. Within this perspective, a sub-ensemble is not a physical entity but a measurement-induced construct, defined by a linear measurement functional acting on the global field and specified by a spatial weighting profile.

In this formulation, correlations between sub-ensemble signals acquire a precise and general meaning, reflecting whether distinct observables sample a common underlying stochastic system [22–25]. In this sense, spin-noise correlations are naturally correlations between measurement-defined sub-ensemble observables. They are determined exclusively by the covariance operator of the underlying spin-fluctuation field, and by the relative orientation of the corresponding measurement functionals within the induced statistical geometry. In particular, statistical independence cannot be inferred from microscopic transport mechanisms or specific experimental arrangement; it is defined instead by orthogonality in the covariance-induced inner-product space. Thus, conventional criteria for sub-ensemble independence based on single-particle diffusion lengths fail to capture the collective, field-theoretic nature of spin fluctuations and are fundamentally incomplete.

Spin fluctuations in thermal vapors are well described by a diffusion–relaxation Ornstein–Uhlenbeck stochastic field, whose statistical properties are fully encoded in the equal-time covariance operator [26,27]. Within this framework, spatial correlations are naturally understood as arise from the collective spectrum of fluctuation modes, rather than from individual atomic motion. Building on this perspective, spatial confinement and boundary conditions act directly on the mode spectrum, suppressing long-wavelength components and thereby imposing an intrinsic bound on the effective number of statistically distinguishable spatial observables. Correlations between spatially resolved probes thus emerge as a spectral inevitability whenever multiple measurements project onto a common, finite-dimensional modal subspace.

The present work develops a unified, experiment-independent theoretical framework for sub-ensemble correlations in diffusion-coupled stochastic fields. While single-cell, multi-channel atomic magnetometry provides a concrete illustrative realization, the formalism is not tied to any specific experimental platform. Within this framework, the notion of spatial sampling overlap is introduced to formalize the statistical breakdown of sub-ensemble independence induced by diffusion-mediated correlations. Spatial resolution is thereby recast as a field-level spectral constraint, governed by the covariance structure of the underlying stochastic dynamics.

## 2. Model

### 2.1 Sub-ensembles as statistical projections

In a diffusion-coupled thermal atomic system, thermal motion continuously redistributes atoms throughout the cell, precluding the existence of dynamically isolated subsystems capable of sustaining independent evolution. Any spatially localized collection of atoms is therefore transient and lacks temporal persistence, and cannot be uniquely associated with a statistically independent component of the collective spin dynamics. As a result, identifying a sub-ensemble with a spatially confined atomic subset does not admit a well-defined statistical meaning in diffusion-coupled systems.

Any notion of a sub-ensemble must therefore be statistical in nature and arise solely from how the system is interrogated. Sub-ensembles are accordingly defined as measurement-induced statistical projections of a global spin-fluctuation field.

From this perspective, although a channel is often implemented experimentally by a localized probe beam, at the statistical level it is more appropriately regarded as a measurement-defined observable specified by its spatial weighting function acting on a common stochastic field. With this understanding, the signal associated with channel-$i$ takes the form of a linear functional in the Hilbert space $L^2(\Omega)$,

$$S_i(\mathbf{r},t) = \int_\Omega W_i(\mathbf{r}) P(\mathbf{r},t) d^3r = \langle W_i(\mathbf{r}), P(\mathbf{r},t) \rangle, \qquad (2.1)$$

where $W_i(\mathbf{r})$ is a real, non-negative spatial weighting function determined by the probe geometry, absorption profile, and detection scheme (e.g., a Gaussian or flat-top beam profile); $P(\mathbf{r},t)$ denotes the continuous spin-polarization field describing the collective spin dynamics. A sub-ensemble is precisely this statistical projection, as schematically illustrated in figure 1.

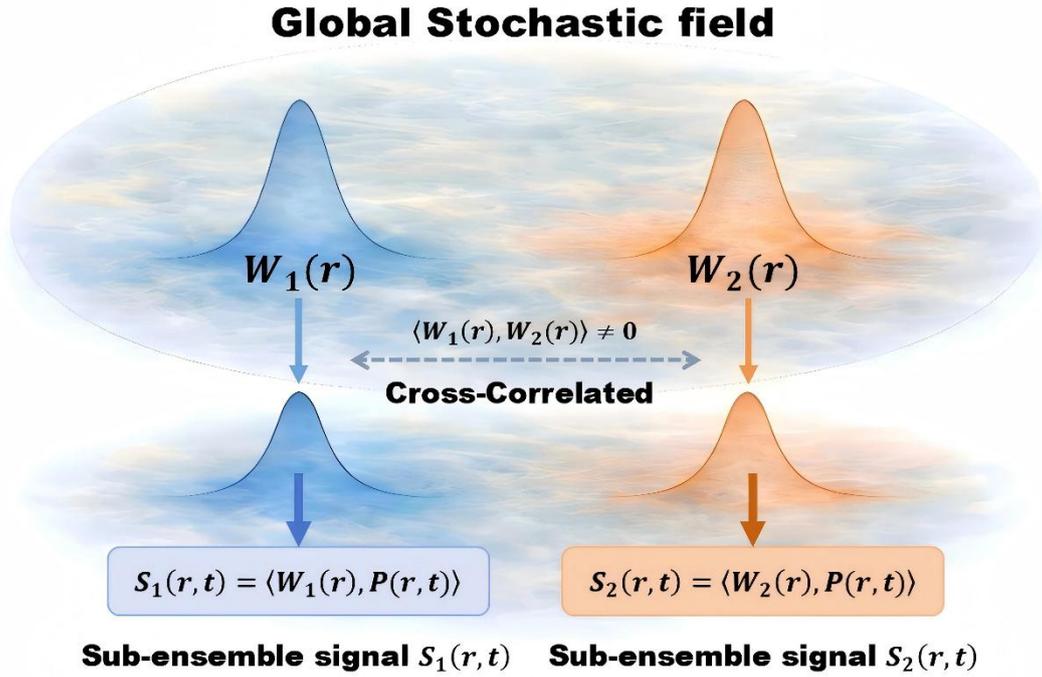

**Figure 1.** Schematic illustration of sub-ensembles as measurement-defined statistical projections of a global stochastic spin field. Distinct measurement weighting functions act on the same underlying field, giving rise to different sub-ensemble observables without implying physically isolated atomic subsystems.

To make the statistical content explicit, we introduce a normalized probability measure encoding the spatial sampling profile of measurement channel-$i$

$$d\mu_i(\mathbf{r}) = \frac{W_i(\mathbf{r})}{\int_\Omega W_i(\mathbf{r}')d^3\mathbf{r}'}d^3\mathbf{r} = \frac{W_i(\mathbf{r})}{Z_i}d^3\mathbf{r}, \qquad (2.2)$$

which allows the sub-ensemble signal to be rewritten as

$$S_i(\mathbf{r},t) = Z_i \int_\Omega P(\mathbf{r},t)d\mu_i(\mathbf{r}). \qquad (2.3)$$

This formulation makes explicit that statistical projection defining a sub-ensemble c corresponds to a weighted spatial average over the global stochastic field. A sub-ensemble is therefore not a physical subset of atoms, but a measurement-defined random variable whose statistics are fixed by the underlying field and the associated weighting function [28]; without a specified weighting, the concept itself is ill-defined.

Within this framework, the statistical relationship between two sub-ensembles $i$ and $j$ is determined by the overlap of their measurement functionals. In the absence of additional correlation structure, they probe disjoint spatial components of the field when their weighting functions have vanishing overlap,

$$\langle W_i, W_j \rangle = \int_\Omega W_i(\mathbf{r}) W_j(\mathbf{r}) d^3 r = 0. \tag{2.4}$$

This condition provides a natural reference point: intuitively, overlapping weighting functions correspond to observables that sample overlapping regions of the underlying field, suggesting the presence of statistical correlations. As will be shown below, diffusion-coupled stochastic dynamics fundamentally modify this intuition through the covariance structure of the field.

**2.2 Spatial correlation structure under diffusion–relaxation dynamics**

In diffusion-dominated thermal atomic vapors, spin fluctuations do not behave as localized, independently evolving perturbations. Instead, they form a collective stochastic field governed by diffusion and relaxation, whose statistical properties extend nonlocally in space. Spatial correlations therefore arise as an inherent feature of the field dynamics and provide the appropriate starting point for a systematic description.

To make this structure explicit, we separate the spin polarization into its fluctuation component,

$$\delta P(\mathbf{r}, t) = P(\mathbf{r}, t) - \langle P(\mathbf{r}, t) \rangle, \tag{2.5}$$

and describe its dynamics with the Bloch–Torrey equation supplemented by a Langevin source [3,29],

$$\partial_t \delta P(\mathbf{r}, t) = D \nabla^2 \delta P(\mathbf{r}, t) - \Gamma \delta P(\mathbf{r}, t) + \xi(\mathbf{r}, t), \tag{2.6}$$

where $D$ is the diffusion coefficient, $\Gamma$ denotes the effective transverse relaxation rate, and $\xi(\mathbf{r}, t)$ represents a zero-mean stochastic source capturing microscopic dissipation at the fluctuation level. Under a standard Markov approximation, the noise is taken to be spatiotemporally δ-correlated,

$$\langle \xi(\mathbf{r}, t) \xi(\mathbf{r}', t') \rangle = 2Q \delta(\mathbf{r} - \mathbf{r}') \delta(t - t'), \tag{2.7}$$

with $Q$ setting the noise strength. This effective description leads to linear, Gaussian stochastic dynamics, thereby defining a diffusion–relaxation Ornstein–Uhlenbeck field [29,30]. The δ-correlated noise assumption should be understood as a coarse-grained description, valid on time and length scales large compared to microscopic collision scales.

To characterize how spin fluctuations decay in space, it is convenient to resolve the field into contributions associated with different spatial scales, motivating a wavevector representation. In this picture, each wavevector $\mathbf{k}$ labels a fluctuation mode associated with a characteristic length scale. Diffusion acts selectively in this basis, preferentially suppresses short-wavelength components while allowing long-wavelength modes to persist.

Owing to the linearity, translational invariance, and Markovian nature of the stochastic dynamics, projecting equation (2.6) onto Fourier space diagonalizes the evolution. Each wavevector mode $\mathbf{k}$ then evolves independently according to a linear Ornstein–Uhlenbeck equation,

$$\frac{d}{dt}\delta P(\mathbf{k},t) = -\gamma_k \delta P(\mathbf{k},t) + \xi(\mathbf{k},t), \tag{2.8}$$

with decay rate $\gamma_k \equiv Dk^2 + \Gamma$, where $k = |\mathbf{k}|$, as derived explicitly in Appendix A.1. In the steady state, the equal-time covariance of each Fourier mode takes the form

$$\tilde{C}_P(\mathbf{k}) = \langle \delta P(\mathbf{k},t)\delta P(-\mathbf{k},t)\rangle = \frac{Q}{\gamma_k} = \frac{Q}{Dk^2 + \Gamma}. \tag{2.9}$$

As visualized in figure 2(a), this equal-time covariance spectrum reveals how diffusion-relaxation operator $\mathcal{L} \equiv -D\nabla^2 + \Gamma$ redistributes statistical weight across spatial scales. High-$k$ (short-wavelength) fluctuations are suppressed as $k^{-2}$, while low-$k$ (long-wavelength) modes overwhelmingly dominate the covariance, and therefore govern long-range correlations in real space. The characteristic scale $\alpha \equiv \sqrt{\Gamma/D}$ does not represent a sharp spectral cutoff or mode elimination, but instead marks a smooth crossover beyond which short-wavelength fluctuations contribute negligibly to the covariance. This redistribution of spectral weight toward low-$k$ modes underlies the emergence of long-range spatial correlations and provides a central organizing principle once genuine mode truncation imposed by finite geometry and boundary conditions is considered later.

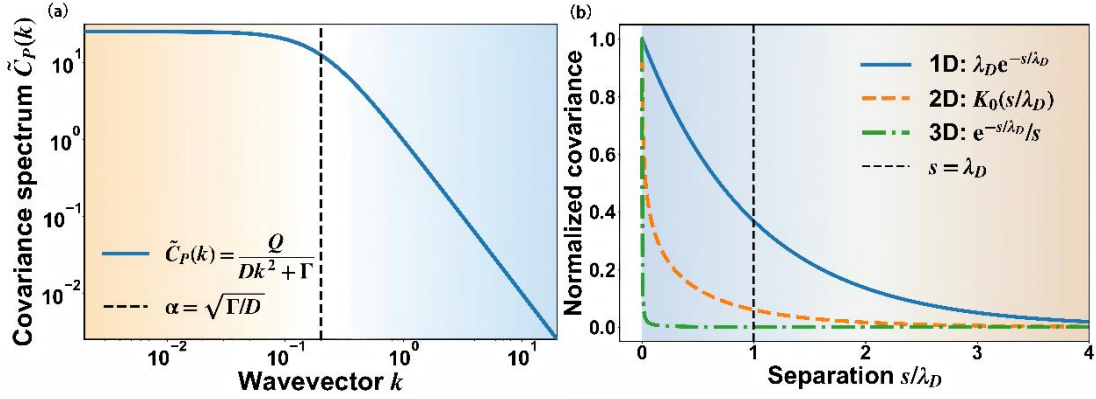

**Figure 2.** Covariance structure in wavevector and real space. (a) Equal-time covariance spectrum $\tilde{C}_P(\mathbf{k})$ in wavevector space, shown on logarithmic axes to emphasize the scale-selective redistribution of statistical weight induced by diffusion and relaxation. The background color gradient encodes scale from long-wavelength (low-$k$, infrared) to short-wavelength (high-$k$, ultraviolet) modes, with the marked scale $\alpha \equiv \sqrt{\Gamma/D}$ indicating the crossover between relaxation- and diffusion-dominated regimes. (b) Bare real-space covariance kernels $C_P(\mathbf{s})$ in one, two, and three spatial dimensions. All curves exhibit a universal exponential decay at large separations $s \gg \lambda_D$, while the short-distance behavior is dimension dependent and reflects non-universal ultraviolet structure. Here 'normalized' refers to rescaling with respect to a finite reference value; the bare covariance kernel is not bounded at short separations. Shaded backgrounds indicate the corresponding short- and long-distance regimes. The explicit dimensional forms of the kernels are derived in Appendix A.3 and are shown here for comparison of their spatial decay behavior rather than absolute magnitude.

Transforming back to real space allows the scale-selective structure encoded in the wavevector spectrum to be expressed directly in terms of spatial correlations. The equal-time spatial covariance is defined as

$$C_P(\mathbf{r}-\mathbf{r}') \equiv \langle \delta P(\mathbf{r},t)\delta P(\mathbf{r}',t) \rangle, \tag{2.10}$$

which, by translational invariance, depends only on the separation $\mathbf{s} = \mathbf{r} - \mathbf{r}'$. Using the variance spectrum $\tilde{C}_P(\mathbf{k})$ of the wavevector modes, the real-space equal-time covariance kernel can be reconstructed as

$$C_P(\mathbf{s}) = \int \frac{d^d k}{(2\pi)^d} e^{i\mathbf{k}\cdot\mathbf{s}} \tilde{C}_P(\mathbf{k}) \equiv Q\mathcal{G}(\mathbf{s}), \tag{2.11}$$

where the kernel $\mathcal{G}(\mathbf{s})$ arises as the Green's function associated with the diffusion–relaxation operator, satisfying $\mathcal{L}\mathcal{G}(\mathbf{s}) = \delta(\mathbf{s})$.

Accordingly, the equal-time spatial covariance itself obeys

$$D\nabla_s^2 C_P(\mathbf{s}) - \Gamma C_P(\mathbf{s}) + Q\delta(\mathbf{s}) = 0, \tag{2.12}$$

as shown by explicit calculation in Appendix A.2. The appearance of the $\delta$-function source term is therefore not an additional assumption, but a direct consequence of the locality of the stochastic forcing. Physically, it reflects the fact that the equal-time covariance quantifies the response of the system to a unit, spatially localized fluctuation injected at zero separation. In this sense, the covariance kernel plays the role of an impulse response of $\mathcal{L}$, encoding how locally generated fluctuations are propagated and attenuated by diffusion and relaxation.

Equation (2.12) is recognized as a modified Helmholtz equation with a point source, which admits standard Yukawa-type Green's kernel [31,32]. The corresponding bare real-space covariance kernels are illustrated in figure 2(b). At large separations, the asymptotic behavior exhibits a universal exponential decay, defining a characteristic decay length

$$\lambda_D \equiv \sqrt{D/\Gamma}. \tag{2.13}$$

Accordingly, $\lambda_D$ characterizes the spatial scale over which correlations persist at large distances, whereas the detailed short-range form of the covariance kernel encodes how local fluctuations are organized on much smaller length scales. The full real-space covariance retains a nontrivial dependence on spatial dimensionality, which, as illustrated in figure 2(b), leaves the asymptotic decay unchanged but gives rise to distinct short-range structures. The dimension-dependent features reflect nonuniversal ultraviolet properties of the continuum description and are collected for completeness in Appendix A.3.

From a field-theoretic perspective, this structure manifests itself through distinct asymptotic regimes of the covariance kernel. At large separations $s \gg \lambda_D$, the covariance kernel decays exponentially, identifying $\lambda_D$ as the spatial correlation length of the spin-noise field; beyond this scale, diffusion-mediated correlations are effectively suppressed by relaxation, and spatially separated fluctuations contribute negligibly to the covariance. In contrast, at short separations $s \ll \lambda_D$, the continuum, δ-correlated noise approximation gives rise to apparent short-range singular behavior. These ultraviolet features do not signal physical divergences, but rather reflect the breakdown of the coarse-grained description at microscopic scales. In realistic systems, they are regularized by finite physical cutoffs, such as the interatomic spacing or the mean free path, ensuring a finite local variance. Moreover, once the field covariance is projected through finite measurement weights, such ultraviolet structure is regularized, a point we return to in the next section. The resulting short-distance structure and its dimensional dependence are analyzed in detail in Appendix A.3.

Taken together, these results establish a clear separation between universal and non-universal features of the diffusion–relaxation covariance kernel, which underlies the robustness of correlation-based constraints and motivates a formulation in which statistical relationships between sub-ensembles are determined at the level of field covariance rather than microscopic detail.

**2.3 Unified functional representation of sub-ensemble correlations**

The bare field covariance fully characterizes the stochastic spin dynamics, but does not by itself determine how correlations appear in experimentally defined observables. Observable correlations arise only after the field is projected through finite measurement weights, which define sub-ensemble signals as spatially averaged random variables. It is therefore natural to formulate correlations directly at the level of these measurement-defined observables.

For a diffusion–relaxation–dominated spin-fluctuation field, the equal-time

covariance between two sub-ensemble signals takes the form

$$\text{Cov}(S_i, S_j) = \langle S_i(t) S_j(t) \rangle = \iint_\Omega W_i(\mathbf{r}) C(\mathbf{r} - \mathbf{r}') W_j(\mathbf{r}') d^3r d^3r'. \qquad (2.14)$$

This expression makes explicit that sub-ensemble correlations are defined through the joint action of the field covariance and the measurement weight functions. This distinction is illustrated schematically in figure 3(a). The apparent short-distance divergence of the kernel itself is absent once correlations are formulated between finite spatial projections.

Importantly, this observation is not pursued as a standalone regularization issue. Rather, it motivates a shift in perspective: sub-ensemble correlations are fundamentally properties of measurement projections acting on a correlated field, rather than attributes of the covariance kernel in isolation. Once correlations are defined through finite spatial weights, the relevant object is no longer the kernel itself, but the bilinear form it induces on the space of measurement functions.

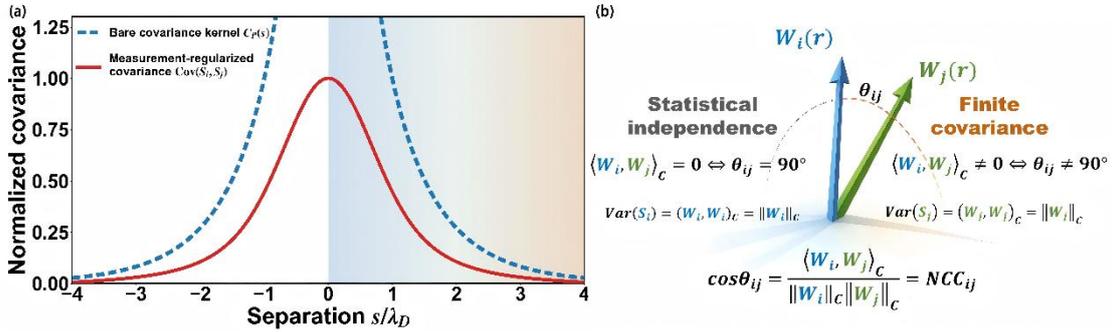

**Figure 3.** Covariance regularization and geometric interpretation of sub-ensemble correlations. (a) Bare and measurement-defined real-space covariances. The bare covariance kernel $C_P(\mathbf{s})$ (dashed) exhibits an apparent short-distance divergence, which is absent once correlations are defined through finite spatial projections (solid). Shading indicates the physically relevant domain $s > 0$. Here 'normalized' refers to rescaling with respect to a finite reference value. (b) Geometric representation of sub-ensemble cross-correlations. Measurement weight functions are represented as vectors in the covariance-induced inner-product space. The angle between vectors encodes the normalized cross-correlation: orthogonality corresponds to statistical independence, while non-orthogonality indicates finite covariance.

This viewpoint naturally leads to a geometric reformulation of sub-ensemble correlations. To make this structure explicit, we introduce the covariance operator $\mathcal{C}$ acting on square-integrable functions,

$$(\mathcal{C}f)(\mathbf{r}) = \int_\Omega C(\mathbf{r}-\mathbf{r}')f(\mathbf{r}')d^3r'. \tag{2.15}$$

which allows the sub-ensemble covariance to be written compactly as

$$\mathrm{Cov}(S_i,S_j) = \langle W_i, \mathcal{C}W_j \rangle. \tag{2.16}$$

This representation makes clear that sub-ensemble cross-correlations are not attributes of individual realizations or trajectories, but arise entirely from the interaction between measurement projections and the covariance structure of the underlying field. It is therefore natural to define a covariance-induced semi-inner product,

$$\langle f,g \rangle_\mathcal{C} \equiv \langle f, \mathcal{C}g \rangle = \iint_\Omega f(\mathbf{r})C(\mathbf{r}-\mathbf{r}')g(\mathbf{r}')d^3rd^3r', \tag{2.17}$$

where the qualifier "semi" reflects the possible presence of fluctuation modes with vanishing variance, which therefore do not contribute to measurable correlations. Within this geometry, the covariance reduces to

$$\mathrm{Cov}(S_i,S_j) = \langle W_i, W_j \rangle_\mathcal{C}. \tag{2.18}$$

A particularly transparent characterization is obtained by introducing the normalized cross-correlation (NCC),

$$NCC_{ij} = \frac{\mathrm{Cov}(S_i,S_j)}{\sqrt{\mathrm{Var}(S_i)\mathrm{Var}(S_j)}} = \frac{\langle W_i, W_j \rangle_\mathcal{C}}{\sqrt{\langle W_i,W_i \rangle_\mathcal{C} \langle W_j,W_j \rangle_\mathcal{C}}}. \tag{2.19}$$

By construction, $NCC_{ij} = 0$ if and only if the corresponding measurement weight functions are orthogonal under the covariance-induced inner product, $\langle W_i, W_j \rangle_\mathcal{C} = 0$. In this case, the two sub-ensembles are statistically independent at the level of measurement-defined observables.

Beyond this formal equivalence, equations (2.17)–(2.19) endow sub-ensemble cross-correlations with a natural geometric structure induced by the field covariance. Within this covariance-induced geometry, weight functions $W_i, W_j$ are represented as vectors, whose norms determine the fluctuation strength of individual channels, while their inner product quantifies the statistical overlap between them. Microscopic stochastic dynamics enter only through the specific realization of the covariance operator $\mathcal{C}$, without altering this $\mathcal{C}$-geometric framework governing sub-ensemble correlations[33,34].

Therefore, normalized cross-correlation admits a direct geometric interpretation as the cosine of the angle between the corresponding measurement vectors, $\cos\theta_{ij}$. Statistical independence corresponds to orthogonality in this $\mathcal{C}$-geometry, while finite cross-correlation reflects non-orthogonal overlap arising from shared sampling of the same fluctuation modes. From this perspective, cross-correlation is elevated from a system-level descriptor to a geometric property of observables: the stochastic field carries an intrinsic, continuous spatial covariance encoded in $C(\mathbf{r}-\mathbf{r}')$, while correlations between sub-ensembles are determined by the relative orientation of the associated measurement projections. This correspondence is illustrated schematically in figure 3(b).

**2.4 Boundary effects in finite domains**

The preceding analysis established sub-ensemble correlations as geometric relations defined by the covariance operator $\mathcal{C}$. In practice, however, this geometry may be realized within a finite spatial domain (e.g. miniaturized and microfabricated vapor cells [35,36]), where boundary conditions restrict the spectrum of accessible fluctuation modes. Finite geometry therefore does not modify the definition of correlations, but reshapes the covariance structure that underlies the

Within a bounded domain $\Omega$, the fluctuation field $\delta P(\mathbf{r},t)$ obeys a linear boundary condition on $\partial\Omega$,

$$\left[\beta_0(\mathbf{r})+\beta_1(\mathbf{r})\partial_n\right]\delta P(\mathbf{r},t)\Big|_{\mathbf{r}=\partial\Omega}=0, \qquad (2.20)$$

which fixes the admissible spectrum of diffusion modes. This unified boundary operator encompasses the Dirichlet ($\beta_0(\mathbf{r})\neq 0, \beta_1(\mathbf{r})=0$), Neumann ($\beta_0(\mathbf{r})=0, \beta_1(\mathbf{r})\neq 0$), and Robin ($\beta_0(\mathbf{r})\neq 0, \beta_1(\mathbf{r})\neq 0$) conditions, parametrizing wall-induced dissipation mechanisms relevant to confined vapor cells [37].

The corresponding linearized Langevin equation,

$$\mathcal{L}\delta P(\mathbf{r},t) = \xi(\mathbf{r},t), \qquad \mathcal{L} \equiv -D\nabla^2 + \Gamma, \tag{2.21}$$

defines a Green's function satisfying

$$\mathcal{L}\mathcal{G}_L(\mathbf{r},\mathbf{r}') = \delta(\mathbf{r}-\mathbf{r}'), \tag{2.22}$$

subject to the same boundary condition on $\mathbf{r} \in \partial\Omega$. Owing to the δ-correlated stochastic forcing, the equal-time covariance can be written as the self-convolution of the Green's function over the finite domain,

$$C_P(\mathbf{r},\mathbf{r}') = \langle \delta P(\mathbf{r})\delta P(\mathbf{r}') \rangle = 2Q\int_\Omega \mathcal{G}_L(\mathbf{r},\mathbf{u})\mathcal{G}_L(\mathbf{r}',\mathbf{u})d^2u. \tag{2.23}$$

In contrast to the infinite-domain case, finite geometry breaks translational invariance, so that the spatial structure of the Green's function is entirely determined by the Laplacian eigenspectrum on $\Omega$, subject to the imposed boundary condition. Introducing the Laplacian eigenmodes $\{\phi_\nu(\mathbf{r})\}$,

$$-\nabla^2\phi_\nu(\mathbf{r}) = k_\nu^2\phi_\nu(\mathbf{r}), \qquad \int_\Omega \phi_\nu(\mathbf{r})\phi_{\nu'}(\mathbf{r})dr^2 = \delta_{\nu\nu'}, \tag{2.24}$$

the Green's function admits the spectral expansion

$$\mathcal{G}_L(\mathbf{r},\mathbf{r}') = \sum_\nu \frac{\phi_\nu(\mathbf{r})\phi_\nu(\mathbf{r}')}{Dk_\nu^2+\Gamma} = \frac{1}{D}\sum_\nu \frac{\phi_\nu(\mathbf{r})\phi_\nu(\mathbf{r}')}{k_\nu^2+\alpha^2}, \qquad \alpha = \sqrt{\Gamma/D}, \tag{2.25}$$

and the equal-time covariance kernel becomes

$$C_P(\mathbf{r},\mathbf{r}') = \frac{2Q}{D^2}\sum_\nu \frac{\phi_\nu(\mathbf{r})\phi_\nu(\mathbf{r}')}{\left(k_\nu^2+\alpha^2\right)^2}. \tag{2.26}$$

This formulation makes explicit that finite-domain effects enter exclusively through the Laplacian spectrum, with different boundary conditions affecting correlations only via their impact on the admissible eigenmodes. Spatial confinement does not continuously deform the covariance kernel; rather, it discretizes the set of admissible diffusion modes and thereby reshapes the statistical support of spatial correlations.

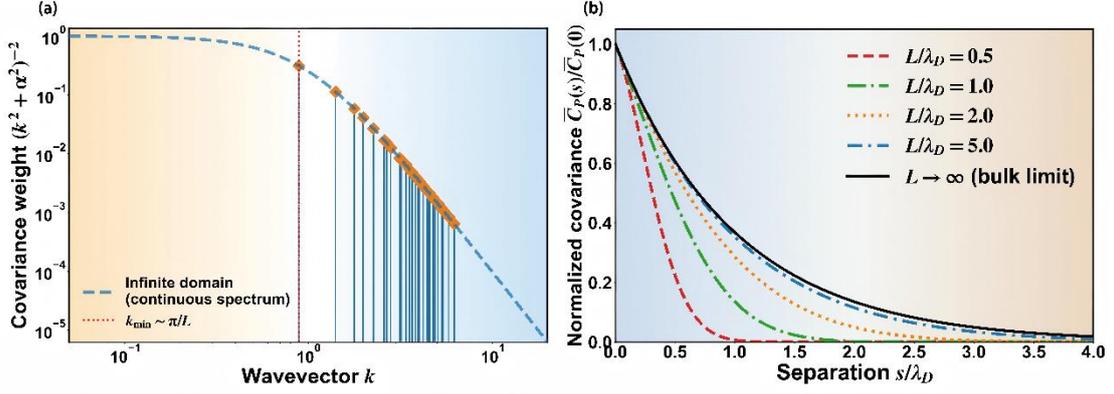

**Figure 4.** Finite-domain covariance in spectral and real space. (a) Covariance weight in wavevector space. The dashed curve compares the continuous spectrum of the infinite domain, while the discrete diamonds indicate the admissible diffusion modes imposed by finite confinement, with the infrared cutoff set by $k_{\min} \sim \pi/L$. (b) Normalized radial covariance $\bar{C}_P(s)/\bar{C}_P(0)$ in real space for representative values of the confinement parameter $L/\lambda_D$. The black solid curve denotes the bulk limit $L \to \infty$. Panels (a) and (b) together demonstrate how finite confinement acts as a structural constraint on the fluctuation spectrum, reshaping spatial correlations through the collective removal of long-wavelength diffusion modes rather than by modifying the underlying decay mechanism.

The underlying mechanism of this spectral restructuring is most transparently revealed in wavevector space. As illustrated in figure 4(a), spatial confinement replaces the continuous covariance spectrum of an unbounded medium with a discrete set of admissible diffusion modes. Finite boundaries therefore do not merely attenuate correlations smoothly; rather, they impose an infrared cutoff that excludes long-wavelength diffusion modes below a boundary-imposed threshold. Finite-domain effects therefore enter as a rigid constraint on the fluctuation modes that govern the steady-state covariance.

To make this mechanism explicit, we consider the Dirichlet boundary condition as a representative case, for which no zero-diffusion mode exists. Physically, this corresponds to uncoated vapor cells with buffer gas, where spin polarization is rapidly lost upon the first wall collision, thereby removing the zero-wavevector mode from the spectrum.

Under the Dirichlet boundary condition,

$$\phi_\nu(\mathbf{r})\big|_{\partial\Omega} = 0, \tag{2.27}$$

the admissible diffusion modes are discretized by the finite domain, with the lowest eigenvalue set by the system size. For a typical square domain $\Omega = [-L/2, L/2]^2$, the orthonormal eigenfunctions takes the form

$$\phi_\nu(\mathbf{r}) = \frac{2}{L}\sin\left(\frac{m\pi}{L}\left(x+\frac{L}{2}\right)\right)\sin\left(\frac{n\pi}{L}\left(y+\frac{L}{2}\right)\right), \quad m,n=1,2,\ldots, \tag{2.28}$$

with eigenmodes

$$k_\nu^2 = \left(\frac{m\pi}{L}\right)^2 + \left(\frac{n\pi}{L}\right)^2, \quad m,n \geq 1. \tag{2.29}$$

The finite system size therefore enforces a nonzero infrared bound on the diffusion spectrum, eliminating longest-wavelength modes from the outset.

The subsequent analysis follows the same spectral-to-real-space logic employed in the unbounded case of section 2.2, with the essential difference that spatial confinement discretizes and truncates the admissible diffusion spectrum.

To elucidate how this spectral restructuring manifests itself in real space, it is convenient to examine a spatial measure of correlations that remains well defined in a confined domain. We therefore consider the radially averaged equal-time covariance

$$\bar{C}_P(s) \equiv \int_\Omega C_P(\mathbf{r},\mathbf{r}+\mathbf{s})d^2r, \quad s = |\mathbf{s}|. \tag{2.30}$$

Figure 4(b) shows the corresponding normalized profiles $\bar{C}_P(s)/\bar{C}_P(0)$ for representative system sizes. Compared to the bulk limit, increasing spatial confinement suppresses long-range correlations in a strongly non-linear manner, with the extended correlation structure collapsing rapidly as low-wavevector diffusion modes are progressively excluded. This behavior reflects not a modification of the underlying diffusion–relaxation dynamics, but the finite-size truncation of the covariance spectrum, which excludes long-wavelength contributions and leaves only short-range fluctuations in the steady state.

To characterize this contraction in a manner independent of overall amplitude, we introduce a second-moment measure of the radially averaged covariance and define an effective correlation half-width. Specifically, we define the zeroth and second moments

$$M_0 = \int_{\mathbb{R}^2} \bar{C}_P(s) d^2s, \qquad M_2 = \int_{\mathbb{R}^2} s^2 \bar{C}_P(s) d^2s. \qquad (2.31)$$

The ratio of these moments provides a statistically well-defined measure of the spatial extent of correlations in confined geometries.

Expressed in a diffusion-mode representation, the second-moment construction leads to a compact expression for the effective correlation half-width,

$$\sigma_P^2 = \frac{1}{2} \frac{\sum_\nu \mu_\nu^{(2)} \left(k_\nu^2 + \alpha^2\right)^{-2}}{\sum_\nu \mu_\nu^{(0)} \left(k_\nu^2 + \alpha^2\right)^{-2}}. \qquad (2.32)$$

Here $\mu_\nu^{(0)}$ and $\mu_\nu^{(2)}$ encodes, respectively, the total covariance weight and the spatially extent of each eigenmode, while the spectral factor $\left(k_\nu^2 + \alpha^2\right)^{-2}$ selects the modes that survive in the steady-state covariance through the competition between diffusion and relaxation.

Although the resulting expression is compact, its explicit evaluation in finite geometries requires careful treatment of modal expansions and convolution structures. For completeness, the full derivation of the second-moment construction is detailed in Appendix B. Importantly, this effective correlation scale encapsulates how finite-domain spectral truncation constrains the spatial support of fluctuations, thereby fixing—independent of measurement implementation—the number of statistically distinguishable spatial observables that can be realized within a single diffusive ensemble.

## 3. Spectral structure and modal capacity of spatial correlations

With the equal-time covariance kernel and finite-domain constraints established, the analysis naturally shifts to a spectral perspective, in which the contraction of spatial correlations is encoded in the restricted set of diffusion modes contributing to the covariance.

Equal-time spatial correlations are fully characterized by the covariance kernel, which induces the aforementioned covariance operator $\mathcal{C}$ defined on the Hilbert space $L^2(\Omega)$. It is self-adjoint, positive, and compact, and therefore admits discrete spectral decomposition

$$\mathcal{C} = \sum_n \gamma_n |\psi_n\rangle\langle\psi_n|, \qquad \gamma_1 \geq \gamma_2 \geq \cdots \to 0, \qquad (3.1)$$

where $\{\psi_n\}$ forms a complete orthonormal basis of spatial covariance modes. This representation makes explicit that equal-time spatial correlations arise from a spectrally weighted superposition of a finite set of dominant covariance modes, rather than from equal contributions of an unbounded family of mutually independent spatial modes. Consequently, the statistical structure supported by a spin-fluctuation field is controlled by the spectral decay of the covariance operator $\mathcal{C}$.

In the infinite-domain or locally homogeneous limit, the covariance is diagonal in wavevector space, admitting a continuous spectral density given in equation (2.9). The spectrum is approximately flat at low wavevectors and decays as $k^{-2}$ at large $k$, reflecting a clear scale selectivity of spatial correlations. Diffusion and relaxation become comparable at

$$k \sim \alpha = \lambda_D^{-1}, \qquad \lambda_D = \sqrt{D/\Gamma} \qquad (3.2)$$

so that only modes with $k \lesssim \alpha$ carry appreciable statistical weight in the covariance.

At the level of the covariance spectrum, this structure defines an effective spectral support from which the number of contributing modes can be estimated. In a $d$-dimensional domain of volume $V$, the density of modes in wavevector space is $V/(2\pi)^d$. Treating modes with $|\mathbf{k}| \lesssim \alpha$ as contributing, the effective number scales as

$$N_{\text{eff}} \simeq \frac{V}{(2\pi)^d} \int_{|\mathbf{k}| \lesssim \alpha} d^d k = C_d \frac{V}{\lambda_D^d}, \qquad C_d = \frac{\Omega_d}{d(2\pi)^d}. \qquad (3.3)$$

where $\Omega_d$ is the surface area of the d-dimensional unit sphere. In this sense, $N_{\text{eff}}$ provides an estimate of the effective rank of the covariance operator $\mathcal{C}$, namely the number of eigenmodes that carry non-negligible statistical spectral weight.

A smooth, cutoff-free alternative is obtained from the spectral participation ratio. In the continuous-spectrum approximation, it takes the form

$$N_{\text{eff}}^{\text{PR}} \equiv \frac{\left(\sum_n \gamma_n\right)^2}{\sum_n \gamma_n^2} \sim \frac{V}{(2\pi)^d} \cdot \frac{\left(\tilde{C}_P(\mathbf{k})d^d k\right)^2}{\int \tilde{C}_P(\mathbf{k})^2 d^d k}. \tag{3.4}$$

which exhibits the same scaling as equation (3.3), namely $N_{\text{eff}} \sim V/\lambda^d$, demonstrating the robustness of the mode-limitation picture with respect to the specific counting prescription.

Finite spatial confinement further reshapes this spectral structure. In a domain of characteristic size $L$, the continuous wavevector spectrum becomes discrete, and a minimum admissible wavevector

$$k_{\min} \equiv \sqrt{\gamma_1(-\nabla^2)} \sim \pi/L \tag{3.5}$$

is imposed by geometry and boundary conditions, where $\gamma_1(-\nabla^2)$ denotes the lowest nonzero Laplacian eigenmode. In general, finite confinement introduces a nonzero infrared cutoff in the spectrum, reducing in the number of contributing modes to

$$N_{\text{eff}} \sim \frac{V}{(2\pi)^d} \frac{\Omega_d}{d}\left(\alpha^d - k_{\min}^d\right), \qquad k_{\min} < \alpha. \tag{3.6}$$

Finite spatial confinement thus reshapes the covariance spectrum by removing its low-wavevector sector, thereby reducing the effective rank of the covariance operator. In particular, for uncoated vapor cells where strong wall depolarization enforces Dirichlet boundary conditions, the zero-wavevector mode is completely excluded, corresponding to a maximal infrared truncation of the spectrum. In the limiting case, when $k_{\min} > \alpha$, the low-$k$ sector is entirely eliminated and $N_{\text{eff}}$ collapses to $O(1)$, indicating the absence of any appreciable long-range spatial correlations.

In summary, the contraction of spatial correlations is not attributable to any specific parameter choice, but an inevitable consequence of the spectral reorganization imposed by finite geometry. In unbounded domains, correlations are sustained by a continuum of low-wavevector modes whose spatial extent is jointly set by diffusion and relaxation. Finite confinement, by contrast, enforces boundary-induced spectral restrictions that introduce an infrared cutoff, selectively eliminating the long-wavelength modes responsible for system-spanning correlations.

Therefore, spatial correlations are not governed by a single correlation length, but by the collective support of a finite set of modes carrying significant spectral weight. As low-wavevector modes are progressively excluded, the number of modes capable of participating in correlated fluctuations is correspondingly reduced; in the extreme limit, long-range spatial correlations vanish altogether, leaving only localized fluctuations.

Finite-domain effects thus impose a fundamental, field-level constraint by reducing the effective rank of the covariance operator. As a result, the number of statistically independent spatial observables is intrinsically limited, irrespective of how probe channels are arranged. Spatial resolution and sub-ensemble independence are therefore governed by the spectral support of the covariance operator, rather than by geometric separation or local correlation lengths.

## 4. Implications for multi-channel detection

Having established a field-theoretic description of sub-ensemble correlations in diffusion-coupled stochastic systems, we now examine its direct implications for multi-channel detection within a single ensemble. Multi-channel atomic magnetometry serves here as a concrete physical realization, not as a source of additional mechanisms, but as a setting in which the geometric and spectral constraints of the covariance operator acquire direct experimental significance.

### 4.1 Spatial sampling overlap as a field-level constraint

To pursue enhanced spatial resolution in ultra-sensitive, noninvasive biomagnetic imaging, such as magnetocardiography (MCG) and magnetoencephalography (MEG) [38–43], recent efforts have increasingly shifted from conventional multi-cell arrays [44–50] toward integrating multiple probe channels within a single vapor cell [12–21]. In this architecture, locally resolved responses are obtained by defining multiple discrete measurement-induced sub-ensembles through distinct optical sampling weights, with the implicit expectation that each channel probes an independent 'virtual cell' [51–53].

In diffusion-dominated vapors, however, this assumption is generically violated. Atomic motion couples nominally distinct sampling volumes through the shared covariance spectrum of the spin-fluctuation field, causing different measurement weights to project onto overlapping fluctuation modes. Under sufficiently dense channel spacing, this overlap grows appreciable, degrading statistical independence and imposing a fundamental limit on the extractable spatial information. Crucially, this limitation is not an artifact of imperfect engineering or microscopic transport details, but follows directly from the covariance structure of the underlying stochastic field.

Conventional single-particle diffusion arguments treat this effect in a largely phenomenological manner as 'crosstalk', inferring channel independence from the typical atomic displacement prior to decoherence [54]. Accordingly, experimental implementations often resort to substantially enlarged inter-channel separations to satisfy an empirically defined 'crosstalk-free' condition [21,55–57]. Such approaches, however, fail to capture the collective constraint identified here: the limiting correlations originate at the field-level and are governed by the spectral structure of the covariance operator, rather than by individual atomic trajectories.

We formalize this constraint as spatial sampling overlap (SSO): the non-orthogonality of measurement-induced sub-ensembles under the covariance-induced geometry of a diffusion-coupled stochastic field. It is a field-level property, reflecting the inevitable project of multiple measurement operators onto overlapping fluctuation modes of a common diffusive ensemble, independent of geometric beam overlap. Channel independence is therefore not an absolute attribute, but a statistical notion whose validity depends on whether SSO is suppressed below a relevant noise floor or tolerance threshold.

### 4.2 Spectral control of spatial sampling overlap

To render SSO quantitatively tractable, the general covariance-operator framework is specialized to the explicit spatial kernel of a spin-fluctuation field. For a two-dimensional spin-fluctuation field, the Green's function of the modified Helmholtz operator admits the analytical form

$$\mathcal{G}_L(\mathbf{r}) = -\frac{1}{2\pi D} K_0(\alpha \mathbf{r}), \tag{4.1}$$

where $K_0$ is the modified Bessel function of the second kind and $\alpha = \sqrt{\Gamma/D}$. Although $\mathcal{G}_L(\mathbf{r})$ admits an analytical Bessel form, its detailed functional shape is immaterial for SSO. Inter-channel correlation is controlled solely by the spatial extent over which a typical fluctuation contributes coherently to multiple measurements, uniquely captured by the second central moment of the covariance kernel. We therefore adopt a second-moment–preserving parametrization to define an effective sampling width.

In multi-channel measurements, spatial correlations of the spin stochastic field reflect two conceptually distinct contributions: diffusion-mediated propagation encoded in the fluctuation covariance, and the finite spatial extent of the measurement weights characterized by their width. The resulting effective kernel, given by the convolution of the covariance kernel with the measurement weights, therefore has a second moment of

$$\langle r^2 \rangle_{\text{tot}} = \langle r^2 \rangle_R + \langle r^2 \rangle_h = w^2 + 8\lambda_D^2, \tag{4.2}$$

Here, $\langle r^2 \rangle_R = w^2$ is the normalized second moment of the sampling weight (assumed Gaussian with characteristic width $w$), while $\langle r^2 \rangle_h = 8\lambda_D^2$ denotes the normalized second moment of the probability kernel associated with fluctuation covariance.

Spatial sampling overlap is controlled by the low-order sector of the covariance spectrum; its geometric characterization therefore collapses to a single effective length scale. Defining an equivalent Gaussian representation $G(r; \sigma_{\text{eff}})$ via $2\sigma_{\text{eff}}^2 = \langle r^2 \rangle_{\text{tot}}$, one obtains the effective sampling half-width

$$\sigma_{\text{eff}}^2 = \frac{w^2}{2} + 4\lambda_D^2. \tag{4.3}$$

This single length scale fully characterizes spatial sampling overlap in multi-channel detection. Accordingly, in diffusion-dominated systems the dominant spatial scale of cross-channel correlations is set by transport, while measurement geometry enters only as a sub-leading correction, consistent with its reduced weight in the effective second moment.

Notably, this Gaussian representation introduced here is not an approximation to the physical covariance kernel, but a second-moment-preserving parametrization that isolates the spatial extent relevant for cross-channel correlations. All scaling relations thus depend solely on the low-order structure of the covariance spectrum and remain insensitive to the detailed kernel shape. The intermediate steps leading to this result are detailed in Appendix C.

Within this representation, the normalized cross-correlation (NCC) between two channels separated by a distance $d$ takes the compact form

$$C(\sigma_{\text{eff}}, d) = \exp\left[-\frac{d^2}{4\sigma_{\text{eff}}^2}\right]. \tag{4.4}$$

providing a direct quantitative measure of spatial sampling overlap governed by the single effective length scale $\sigma_{\text{eff}}$.

We next illustrate the implications of the SSO model in a representative spin-exchange–relaxation-free (SERF) regime [54], with a characteristic correlation length $\lambda_D = 0.25\,\text{mm}$, for a concrete measurement geometry. Figures 5(a)-(d) visualize the spatial structure of the combined effective sampling response

$$\Phi_{\text{tot}}(\mathbf{r}) = \mathcal{C}W_1(\mathbf{r}) + \mathcal{C}W_2(\mathbf{r}) = \sum_n \gamma_n \psi_n(\mathbf{r})\left(\langle\psi_n|W_1\rangle + \langle\psi_n|W_2\rangle\right), \tag{4.5}$$

which represents the joint spatial weighting of covariance modes simultaneously sampled by two identical channels separated by $d$. After normalization, the heat maps indicate where in space covariance-mode content is jointly accessed by both channels, rather than any absolute fluctuation amplitude. They should not be interpreted as optical beam profiles or steady-state polarization distributions. As $d$ decreases, the two measurement operators transition from sampling largely disjoint combinations of covariance modes to coherently projecting onto the same set of dominant low-order modes, marking a crossover from geometrical separability to modal indistinguishability.

This crossover is quantitatively captured in figure 5(e) by the collapse of the two-peak structure in the $y=0$ cross-sectional profiles. In panels (a)–(e), the individual sampling weights are Gaussian with a fixed $1/e^2$ diameter of $1.0\,\text{mm}$. Figure 5(f) shows how SSO reorganizes as a function of channel separation, demonstrating that modal non-independence persists over a finite range set by $\sigma_{\text{eff}}$ and is only weakly mitigated by reducing the sampling radius. By mapping SSO into the $(\sigma_{\text{eff}}, d)$ plane, figure 5(g) exposes a fundamental boundary separating geometrically tunable regimes from diffusion-limited modal indistinguishability. For any prescribed correlation threshold, this map directly identifies the admissible parameter region, corresponding to values above the chosen contour, as further discussed in section 4.3.

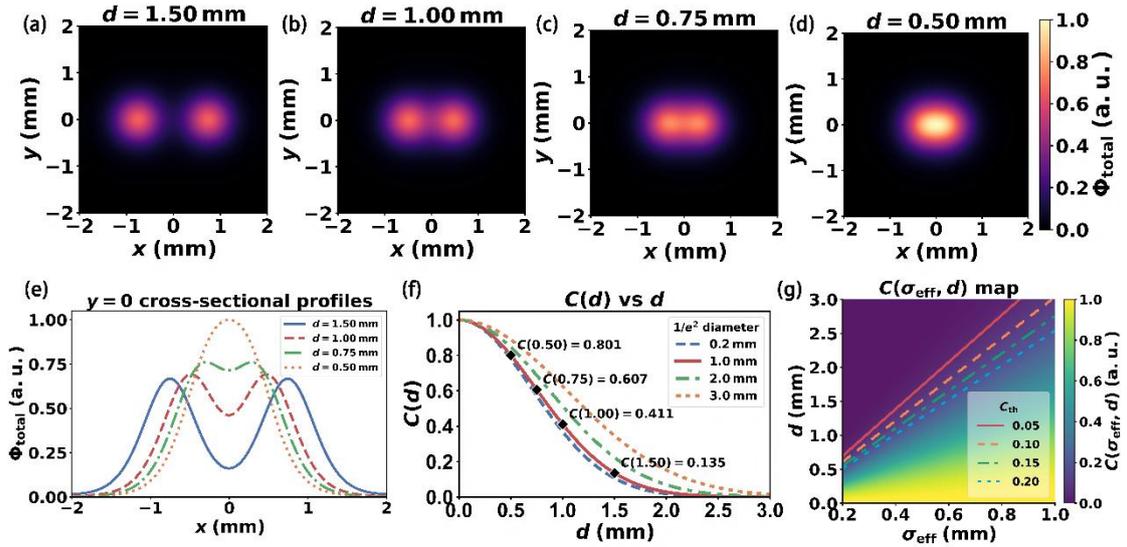

**Figure 5.** Spatial sampling overlap as a consequence of covariance-induced modal compression in a diffusion-dominated stochastic field. (a–d) Two-dimensional maps of the combined effective sampling response $\Phi_{\text{tot}}(\mathbf{r})$ shown for channel separations $d = 1.5,\ 1.0,\ 0.75$ and $0.5\,\text{mm}$, respectively, with a fixed Gaussian diameter of $1/e^2 = 1.0\,\text{mm}$. These maps visualize the joint spatial weighting of covariance modes simultaneously sampled by both channels and are normalized to the global maximum; not the origin of correlations. (e) Normalized modal cross-sections of the effective sampling response in (a–d) taken along $y = 0$, illustrating the progressive collapse of two distinct modal supports into a single low-order modal structure as the channel separation is reduced.

(f) Normalized cross-correlation $C(d)$ as a function of channel separation four different measurement diameters ($1/e^2 = 0.2, 1.0, 2.0$ and $3.0\text{mm}$), highlighting the weak dependence on beam size in the diffusion-dominated regime. Black diamonds indicate the cases in (a–d). (g) Two-parameter map of the normalized sampling overlap in the $(\sigma_{\text{eff}}, d)$ plane, delineating the boundary between regimes of modal distinguishability and unavoidable modal overlap, and identifying the admissible parameter region above a prescribed correlation threshold.

**4.3 Engineering limit on achievable spatial resolution**

In practical multi-channel implementations, spatial sampling overlap manifests operationally as cross-channel leakage whose impact is set by its magnitude relative to the intrinsic noise floor. A convenient and conservative engineering criterion is therefore to constrain the magnitude of leakage-induced contributions to remain within a prescribed fraction of the noise floor.

To this end, we decompose the output of channel $i$ into three contributions,

$$S_i(t) = S_i^{(\text{self})}(t) + S_{i \leftarrow j}^{(\text{leak})}(t) + n_i(t), \tag{4.6}$$

where $S_i^{(\text{self})}$ denotes the desired response associated with the local sampling weight, $S_{i \leftarrow j}^{(\text{leak})}(t)$ arises from diffusion-mediated overlap with a neighboring channel $j$, and $n_i$ represents additive measurement noise with deviation $\sigma_i := \sqrt{\text{Var}(n_i)}$.

Both the desired signal and the leakage term are linear functionals of the same underlying stochastic spin-fluctuation field and therefore share an identical second-order statistical structure, with their relative magnitude controlled by NCC. In particular, a conservative upper bound on leakage amplitude, determined by the second-order statistics of the field, is therefore given by

$$\left|S_{i \leftarrow j}^{(\text{leak})}\right| \leq C(d) \cdot \left|S_i^{(\text{self})}\right| \tag{4.7}$$

where $C(d)$ is the NCC between channels separated by a distance $d$.

Introducing the signal-to-noise ratio of channel $i$, $\text{SNR}_i = \left|S_i^{(\text{self})}\right|/\sigma_i$, and an admissible tolerance $0 < \varepsilon < 1$ that sets the maximum leakage-to-noise ratio,

$\left|S_{i \leftarrow j}^{(\text{leak})}\right| \leq \varepsilon \sigma_i$, the requirement for acceptable channel independence reduces to

$$C(d) \leq \frac{\varepsilon}{\text{SNR}_i} = C_{th}. \tag{4.8}$$

Substituting the analytical form of the NCC obtained in equation (4.4) then yields a minimum channel separation,

$$d_{crit} \geq 2\sigma_{\text{eff}} \sqrt{\ln\left(\frac{\text{SNR}_i}{\varepsilon}\right)}. \tag{4.9}$$

Equation (4.9) establishes a diffusion-imposed upper bound on the achievable spatial resolution in multi-channel detection: for a given SNR and tolerated level of residual correlation, resolving independent measurement-defined sub-ensembles requires spatial separations exceeding a characteristic scale set by $\sigma_{\text{eff}}$. Crucially, this resolution limit cannot be relaxed by geometric optimization—whether through beam shaping, channel spacing, or array layout—because it originates from the infrared structure of the fluctuation covariance spectrum, fixed by diffusion, relaxation, and boundary-induced spectral confinement.

From this perspective, the ultimate limit to spatial resolution in diffusion-coupled systems is not set by optical point-spread considerations, but by a field-level statistical constraint determined by the spatial extent over which spin fluctuations remain correlated. Engineering criteria for multi-channel architectures must therefore be formulated at the level of stochastic-field covariance, rather than inferred from purely optical resolution arguments.

## 5. Discussion

In diffusion–relaxation stochastic fields, the number of statistically distinguishable spatial modes is intrinsically bounded by the covariance spectrum, independent of measurement realization. Sub-ensembles, as statistical projections of a global fluctuation field, are therefore limited by the effective rank of the covariance operator, rather than by spatial channel density. Beyond this spectral capacity, adding further measurement channels does not increase the accessible independent information.

From this perspective, correlations between sub-ensembles emerge as a spectral inevitability. They are collectively supported by a finite set of low-wavevector modes carrying appreciable spectral weight, rather than being determined by single-particle transport trajectories or microscopic exchange paths. Distinct sub-ensembles arise as different measurement projections acting on a common stochastic field; whenever these projections are non-orthogonal in the covariance-induced geometry, they necessarily share fluctuation modes and are therefore statistically correlated. This viewpoint clarifies a common implicit assumption: geometric separation alone does not guarantee statistical independence [58]; whenever distinct channels probe overlapping sectors of the covariance spectrum, their outputs are unavoidably correlated.

In finite-sized vapor cells, such as miniaturized or microfabricated geometries [59–62], the limitation of sub-ensemble statistical independence is driven not merely by enhanced diffusion-mediated coupling, but by a systematic spectral reorganization imposed by boundary conditions. Finite confinement imposes an infrared cutoff that suppresses, or entirely eliminates, the low-wavevector modes that sustain long-range correlations. As a result, the effective rank of the covariance operator is compressed, forcing multiple sub-ensembles to project onto the same finite-dimensional modal subspace and rendering them statistically indistinguishable. In this regime, increasing the number of spatially distinct probes does not yield proportional growth in statistically independent content. Instead, the accessible spatial information saturates as the finite set of supported fluctuation modes is exhausted, imposing an intrinsic bound on spatial multiplexing in diffusion-coupled systems.

Within the present framework, spatial resolution in a multi-sub-ensemble system acquires a purely statistical meaning: it is the scale at which distinct measurement-defined sub-ensembles cease to be statistically distinguishable under a prescribed noise tolerance. Equivalently, spatial resolution is set by the extent to which measurements access independent fluctuation modes of the underlying field; once covariance-induced correlations become appreciable, geometrically separated probes no longer yield independent local responses, and spatial structure becomes statistically unresolved.

This diffusion-mediated loss of statistical resolvability is captured by the notion of spatial sampling overlap.

The results presented here rely only on a minimal set of assumptions: a diffusive stochastic field with linear relaxation, linear functional measurements, and a statistical description at the level of equal-time covariances. Within this regime, the covariance spectrum, the effective modal content, and the resulting statistical bounds are universal. Strong nonlinear feedback, pronounced non-Markovian noise, or nonequilibrium driving may qualitatively modify sub-ensemble statistics by introducing memory effects and irreversible probability flows that reshape correlation structure beyond the Markovian, covariance-based description [63–68]. Recent demonstrations of reservoir-engineered nonreciprocity, dynamically generated inter-channel quantum correlations, and concurrent spin–light squeezing in hot atomic ensembles exemplify regimes in which correlations are actively generated by coherent or nonequilibrium dynamics, rather than being kinematically constrained by diffusion alone [69–72].

Although single-cell multi-channel atomic magnetometry serves as a concrete demonstration, the conclusions are not tied to any specific experimental platform. Related diffusive fluctuation systems, such as noise fields encountered in biomagnetic imaging and fluctuation-based imaging schemes, are naturally described at the level of spatially correlated stochastic fields with covariance-structured observables. Within such settings, analogous covariance-spectral constraints are expected to govern statistical distinguishability. Systematic exploration of these generalizations constitutes a natural direction for future work.

Ultimately, this work reframes sub-ensemble correlations beyond geometric intuition and single-particle transport pictures, casting them as an intrinsically field-level statistical problem governed by the spectral structure of the covariance operator. At its core, this work formulates a unified theoretical language linking spatial modes, covariance spectra, and statistical distinguishability, in which sub-ensembles are not physical subsystems but statistical constructs whose independence is bounded by the finite set of accessible fluctuation modes.

## Data availability statement

All data that support the findings of this study are included within the article (and any supplementary files).

## Acknowledgment

This work was supported by the National Natural Science Foundation of China for Excellent Young Scientist (Overseas) (Grant No.37110101); the National Natural Science Foundation of China (Grant No.77051001); the Innovation Program for Quantum Science and Technology (Grant No.2021ZD0300503); and the Quantum Science and Technology-National Science and Technology Major Project (Grant No.2021ZD0300403).

# Appendix A. Derivation of spin-fluctuation covariance kernel

## A.1. Spectral decomposition into independent Ornstein–Uhlenbeck modes

The purpose of Appendix A.1 is not merely to reproduce intermediate algebraic steps, but to make explicit the structural reason why the fluctuation spectrum appearing in equations (2.8)–(2.9) is both diagonal in wavevector space and governed by a single relaxation rate $\gamma_k \equiv Dk^2 + \Gamma$.

We consider the fluctuation field $\delta P(\mathbf{r},t)$ governed by the linear stochastic diffusion–relaxation dynamics introduced in section 2.2. Owing to the linearity and translational invariance of the underlying operator, the fluctuation generator $\mathcal{L} \equiv -D\nabla^2 + \Gamma$ admits a complete set of plane-wave eigenfunctions. Consequently, the natural variables in which both the dynamics and the noise statistics diagonalize are the spatial Fourier modes $\delta P(\mathbf{k},t)$.

We adopt the following $d$-dimensional spatial Fourier transform convention:

$$\delta P(\mathbf{k},t) = \int d^d r\, e^{-i\mathbf{k}\cdot\mathbf{r}} \delta P(\mathbf{r},t), \tag{A.1}$$

with the inverse transform implicitly defined by

$$\delta P(\mathbf{r},t) = \int \frac{d^d k}{(2\pi)^d} e^{i\mathbf{k}\cdot\mathbf{r}} \delta P(\mathbf{k},t). \tag{A.2}$$

Under this convention, the Fourier representation of the spatial $\delta$-function reads

$$\int d^d r\, e^{-i(\mathbf{k}+\mathbf{k}')\cdot\mathbf{r}} = (2\pi)^d \delta(\mathbf{k}+\mathbf{k}'). \tag{A.3}$$

Projecting the Langevin equation (2.6) onto this eigenbasis yields a decoupled evolution equation for each wavevector mode:

$$\frac{d}{dt}\delta P(\mathbf{k},t) = -\gamma_k \delta P(\mathbf{k},t) + \xi(\mathbf{k},t), \qquad \gamma_k = Dk^2 + \Gamma, \tag{A.4}$$

where $k \equiv |\mathbf{k}|$. Equation (A.4) corresponds to equation (2.8) in the main text. It shows that each Fourier component evolves independently as a linear Ornstein–Uhlenbeck process, with a decay rate set by the spectrum of the diffusion–relaxation operator.

Crucially, this Ornstein–Uhlenbeck structure is not an additional modeling assumption, but a direct consequence of linearity, translational invariance, and Markovian noise. No further assumptions beyond those already stated in section 2.2 are required. The stochastic Langevin forcing $\xi(\mathbf{r},t)$, assumed white in space and time, remains diagonal under the same spectral decomposition, leading to

$$\langle \xi(\mathbf{k},t)\xi(\mathbf{k}',t') \rangle = 2Q(2\pi)^d \delta(\mathbf{k}+\mathbf{k}')\delta(t-t'). \tag{A.5}$$

The statistical properties of the fluctuation field therefore reduce to those of a family of independent Ornstein–Uhlenbeck modes. In the stationary limit, each mode possesses a Lorentzian power spectral density,

$$S_P(\mathbf{k},\omega) = \int_{-\infty}^{+\infty} dt\, e^{i\omega t} \langle \delta P(\mathbf{k},t) P(-\mathbf{k},0) \rangle = \frac{2Q}{\omega^2 + \gamma_k^2}, \tag{A.6}$$

from which the equal-time variance follows directly,

$$\tilde{C}_P(\mathbf{k}) \equiv \langle |\delta P(\mathbf{k},t)|^2 \rangle = \int \frac{d\omega}{2\pi} S_P(\mathbf{k},\omega) = \frac{Q}{\gamma_k} = \frac{Q}{Dk^2 + \Gamma}. \tag{A.7}$$

Equation (A.7) corresponds to equation (2.9) in the main text and provides the momentum-space covariance of the spin-fluctuation field. Its simple rational form reflects the resolvent of the generator $\mathcal{L}$ and underlies the emergence of a modified Helmholtz equation for the real-space covariance kernel, as discussed in Appendix A.2.

### A.2. From spectral covariance to the modified Helmholtz equation

In Appendix A.2 we establish explicitly how the momentum-space covariance obtained in Appendix A.1 leads to a modified Helmholtz equation for the equal-time spatial covariance kernel. The derivation is purely mathematical and serves to justify the transition from equation (2.11) to equation (2.12) in the main text.

Starting from the steady-state variance spectrum of the fluctuation modes in equation (A.7), we define the equal-time spatial covariance kernel in real space as the inverse Fourier transform

$$C_P(\mathbf{s}) \equiv \langle \delta P_z(\mathbf{r},t) \delta P_z(\mathbf{r}',t) \rangle, \quad \mathbf{s} = \mathbf{r} - \mathbf{r}', \tag{A.8}$$

which, by translational invariance, depends only on the displacement $\mathbf{s}$.

Using the Fourier convention introduced in Appendix A.1, the covariance kernel

can be written as

$$C_P(\mathbf{s}) = \int \frac{d^d k}{(2\pi)^d} e^{i\mathbf{k}\cdot\mathbf{s}} \tilde{C}_P(\mathbf{k}) = Q \int \frac{d^d k}{(2\pi)^d} \frac{e^{i\mathbf{k}\cdot\mathbf{s}}}{\gamma_k} \equiv QG(\mathbf{s}), \quad (A.9)$$

where we have introduced the kernel $G(\mathbf{s})$ for notational convenience. Equation (A.9) corresponds to equation (2.9) in the main text.

The integral kernel $G(\mathbf{s})$ is recognized as the Green's function of the diffusion–relaxation operator $\mathcal{L} \equiv -D\nabla^2 + \Gamma$, in the sense that it satisfies

$$\mathcal{L}G(\mathbf{s}) = \delta(\mathbf{s}). \quad (A.10)$$

This relation follows directly by applying the operator $-D\nabla_s^2 + \Gamma$ to the Fourier representation of $G(\mathbf{s})$.

To make this explicit, we act with the Laplacian on the covariance kernel $C_P(\mathbf{s})$

$$\nabla_s^2 C_P(\mathbf{s}) = \int \frac{d^d k}{(2\pi)^d} e^{i\mathbf{k}\cdot\mathbf{s}} (-k^2) \tilde{C}_P(\mathbf{k}), \quad (A.11)$$

and therefore

$$-D\nabla_s^2 C_P(\mathbf{s}) = \int \frac{d^d k}{(2\pi)^d} e^{i\mathbf{k}\cdot\mathbf{s}} (Dk^2) \tilde{C}_P(\mathbf{k}). \quad (A.12)$$

Similarly, the relaxation term gives

$$\Gamma C_P(\mathbf{s}) = \int \frac{d^d k}{(2\pi)^d} e^{i\mathbf{k}\cdot\mathbf{s}} (\Gamma) \tilde{C}_P(\mathbf{k}). \quad (A.13)$$

Adding equations (A.12) and (A.13), we obtain

$$(-D\nabla_s^2 + \Gamma) C_P(\mathbf{s}) = \int \frac{d^d k}{(2\pi)^d} e^{i\mathbf{k}\cdot\mathbf{s}} \left[ (Dk^2 + \Gamma) \tilde{C}_P(\mathbf{k}) \right]. \quad (A.14)$$

Substituting the explicit form of $\tilde{C}_P(\mathbf{k})$ from Equation (A.7), the right-hand side of Equation (A.14) reduces to

$$Q \int \frac{d^d k}{(2\pi)^d} e^{i\mathbf{k}\cdot\mathbf{s}} = Q\delta(\mathbf{s}). \quad (A.15)$$

We thus arrive at

$$D\nabla_s^2 C_P(\mathbf{s}) - \Gamma C_P(\mathbf{s}) + Q\delta(\mathbf{s}) = 0. \tag{A.16}$$

which is precisely equation (2.12) in the main text.

### A.3. Dimensional structure of the diffusion–relaxation covariance kernel

While the explicit functional form of the covariance kernel depends on spatial dimensionality, this dependence is not merely technical. Rather, it reflects how diffusive fluctuations distribute spectral weight across length scales in different dimensions.

The purpose of this Appendix is therefore not to introduce new physics, but to make explicit how a single, dimension-independent correlation length coexists with dimension-dependent short-range structures in the real-space covariance.

For completeness, we record the explicit real-space representation of the equal-time covariance kernel, which can be written in the unified form

$$C_P(\mathbf{s}) = \int \frac{d^d k}{(2\pi)^d} e^{i\mathbf{k}\cdot\mathbf{s}} \frac{Q}{D} \frac{1}{k^2 + \alpha^2} = \frac{Q}{D} I_d(s), \qquad \alpha \equiv \lambda_D^{-1}, \tag{A.17}$$

with

$$I_d(\mathbf{s}) \equiv \int \frac{d^d k}{(2\pi)^d} \frac{e^{i\mathbf{k}\cdot\mathbf{s}}}{k^2 + \alpha^2}. \tag{A.18}$$

The dimension-dependent structure of the covariance kernel is therefore fully encoded in the standard $d$-dimensional Fourier integral $I_d(\mathbf{s})$, whose explicit form depends on spatial dimensionality.

### A.3.1. Three dimensions ($d$=3)

In three spatial dimensions, the integral in equation (A.18) can be evaluated in closed form, yielding

$$I_3(s) = \frac{1}{4\pi s} e^{-\alpha s}. \tag{A.19}$$

Accordingly, the real-space covariance kernel reduces to the familiar Yukawa form,

$$C_P^{(3D)}(s) = \frac{Q}{4\pi D} \frac{e^{-s/\lambda_D}}{s}, \qquad s = |\mathbf{r} - \mathbf{r}'|. \tag{A.20}$$

At short separations $s \ll \lambda_D$, the kernel exhibits an apparent Coulomb-like singularity $\sim 1/s$, reflecting the ultraviolet sensitivity of the continuum description under the assumption of spatially δ-correlated noise. In physical systems, this divergence is regularized by microscopic cutoffs such as the interatomic spacing or the mean free path. At large separations $s \gg \lambda_D$, the covariance decays exponentially, with the characteristic length scale $\lambda_D$ controlling the spatial extent of correlations.

Importantly, such Coulomb-like divergence does not indicate a physical pathology, but rather encodes the fact that the continuum, δ-correlated noise model assigns equal weight to arbitrarily high wavevector modes. In this sense, the short-distance singularity should be interpreted as a controlled ultraviolet feature of the coarse-grained field theory, rather than a failure of the underlying physical description.

**A.3.2. Two dimensions (*d*=2)**

In two dimensions, the integral $I_2(s)$ is expressed in terms of the modified Bessel function of the second kind,

$$I_2(s) = \frac{1}{2\pi} K_0(\alpha s), \quad (A.21)$$

leading to

$$C_P^{(2D)}(s) = \frac{Q}{2\pi D} K_0(s/\lambda_D). \quad (A.22)$$

In this case, the covariance kernel displays a logarithmic divergence as $s \to 0$, signaling the breakdown of the continuum approximation at short distances, while retaining an exponentially suppressed tail for $s \gg \lambda_D$.

The logarithmic behavior in two dimensions reflects the marginal nature of diffusion at the level of spatial correlations, where neither infrared dominance nor ultraviolet suppression fully prevails, placing two dimensions at the boundary between power-law and finite short-distance behavior.

**A.3.3. One dimension (*d*=1)**

In one dimension, the integral in equation (A.18) yields

$$I_1(s) = \frac{1}{2\kappa} e^{-\alpha|s|}, \quad (A.23)$$

So that

$$C_P^{(1D)}(s) = \frac{Q}{2\alpha D}e^{-\alpha|s|} = \frac{Q\lambda_D}{2D}e^{-|s|/\lambda_D} \tag{A.24}$$

Unlike in higher dimensions, the one-dimensional covariance remains finite at zero separation, reflecting the absence of ultraviolet divergences and dominance of long-wavelength diffusive modes, which suppresses short-wavelength contributions to the equal-time covariance.

## Appendix B. Second-moment characterization of finite-domain covariance kernels

In this Appendix, we present the complete derivation of the second-moment construction used in section 2.4, explicitly retaining finite-domain effects and boundary-induced mode discretization throughout.

In a finite domain $\Omega$, the equal-time covariance kernel $C_P(\mathbf{r}, \mathbf{r}')$ generally depends on both spatial arguments independently. To construct a scalar measure of correlation range, we first eliminate the explicit dependence on the reference position by performing a spatial average over $\mathbf{r}$, defining the radially averaged covariance

$$\bar{C}_P(s) \equiv \int_\Omega C_P(\mathbf{r}, \mathbf{r}+\mathbf{s})d^2r, \qquad s = |\mathbf{s}|. \tag{B.1}$$

This averaging procedure isolates the dependence on the separation $s$ alone, while preserving the full influence of finite geometry and boundary conditions through the integration domain.

In a bounded domain, the covariance kernel admits an expansion in the eigenmodes of the diffusion–relaxation operator,

$$C_P(\mathbf{r}, \mathbf{r}') = \frac{2Q}{D^2}\sum_\nu \frac{\phi_\nu(\mathbf{r})\phi_\nu(\mathbf{r}')}{\left(k_\nu^2 + \alpha^2\right)^2}. \tag{B.2}$$

Substituting equation (B.2) into equation (B.1) yields

$$\bar{C}_P(s) = \frac{2Q}{D^2}\sum_\nu \frac{1}{\left(k_\nu^2 + \alpha^2\right)^2}\int_\Omega \phi_\nu(\mathbf{r})\phi_\nu(\mathbf{r}+\mathbf{s})d^2r, \tag{B.3}$$

The overlap integrals encode the spatial structure of each diffusion mode and can be evaluated analytically for simple geometries or numerically in general.

To make the second-moment construction explicit, it is convenient to separate the shape of the covariance kernel from its overall normalization. We therefore introduce the normalized Green's kernel associated with the Helmholtz operator,

$$g(r) \equiv \frac{\mathcal{G}_L(r)}{\int_\Omega \mathcal{G}_L(r)} = \frac{\mathcal{G}_L(r)}{G_L}, \tag{B.4}$$

which may be interpreted as a probability density over space. The corresponding normalized covariance shape kernel is then defined as the self-convolution

$$h(s) \equiv (g * g)(s) = \int_\Omega g(\mathbf{u}) g(\mathbf{u}+\mathbf{s}) d^2 u, \tag{B.5}$$

which depends only on the relative separation $s$. With these definitions, the equal-time covariance kernel can be written in the factorized form

$$\bar{C}_P(\mathbf{s}) = 2Q(\mathcal{G}_L * \mathcal{G}_L)(\mathbf{s}) = 2Q G_L^2 h(s) = A h(s), \tag{B.6}$$

where the amplitude $A = 2Q G_L^2$ carries all information about the fluctuation strength, while the spatial structure is entirely encoded in $h(s)$.

To characterize the spatial extent of correlations, we define the zeroth and second moments of the radially averaged covariance as

$$M_0 = \int_{\mathbb{R}^2} \bar{C}_P(s) d^2 s, \qquad M_2 = \int_{\mathbb{R}^2} s^2 \bar{C}_P(s) d^2 s. \tag{B.7}$$

From a field-theoretic standpoint, $M_0$ fixes the normalization of the covariance kernel and measures the total variance carried by fluctuations, whereas $M_2$ captures how this variance is distributed in space through its quadratic weighting in separation. Rather than relying on a single analytical correlation length, this second-moment construction provides a statistically well-defined and domain-robust measure of the spatial extent of correlations in confined geometries.

These moments admit an alternative representation obtained by exchanging the order of integration,

$$M_0 = \int_{\mathbb{R}^2} \left( \int_\Omega C_P(\mathbf{r},\mathbf{r}+\mathbf{s}) d^2 r \right) d^2 s = \int_\Omega d^2 r \int_{\mathbb{R}^2} C_P(\mathbf{r},\mathbf{r}+\mathbf{s}) d^2 s, \tag{B.8}$$

$$M_2 = \int_{\mathbb{R}^2} s^2 \left( \int_\Omega C_P(\mathbf{r},\mathbf{r}+\mathbf{s}) d^2 r \right) d^2 s = \int_\Omega d^2 r \int_{\mathbb{R}^2} s^2 C_P(\mathbf{r},\mathbf{r}+\mathbf{s}) d^2 s. \tag{B.9}$$

If the covariance kernel is translationally invariant in the bulk,

$$C_P(\mathbf{r},\mathbf{r}+\mathbf{s}) = C_P(\mathbf{s}) = Ah(s), \tag{B.10}$$

the dependence on the reference position $\mathbf{r}$ drops out, and the integrals reduce to

$$M_0 = \int_\Omega d^2r \int_{\mathbb{R}^2} Ah(s) d^2s = A|\Omega| \int_{\mathbb{R}^2} h(s) d^2s = A|\Omega|, \tag{B.11}$$

$$M_2 = \int_\Omega d^2r \int_{\mathbb{R}^2} s^2 Ah(s) d^2s = A|\Omega| \int_{\mathbb{R}^2} s^2 h(s) d^2s. \tag{B.12}$$

Taking the ratio, the overall amplitude $A$ and the system area $|\Omega|$ cancel identically,

$$\frac{M_2}{M_0} = \frac{A|\Omega| \int_{\mathbb{R}^2} s^2 h(s) d^2s}{A|\Omega|} = \int_{\mathbb{R}^2} s^2 h(s) d^2s = \langle s^2 \rangle_h, \tag{B.13}$$

showing that the second-moment ratio is entirely amplitude-independent and reduces to the variance of the normalized shape function $h(s)$.

To make the second-moment characterization operational in a finite domain, it is instructive to express the effective correlation half-width in a modal representation, defined as

$$2\sigma_P^2 = \langle s^2 \rangle_h = \frac{M_2}{M_0} \Rightarrow \sigma_P^2 = \frac{1}{2} \frac{M_2}{M_0}. \tag{B.14}$$

To make the second-moment characterization explicit in a finite domain, where the covariance kernel admits a discrete spectral representation, we now express the effective correlation half-width in terms of the eigenmodes of the Helmholtz operator. Substituting the modal expansion equation (B.3) into equations (B.7)–(B.9) yields

$$M_0 = \frac{2Q}{D^2} \sum_\nu \frac{\mu_\nu^{(0)}}{(k_\nu^2 + \alpha^2)^2}, \quad \mu_\nu^{(0)} = \int_{\mathbb{R}^2} \left( \int_\Omega \phi_\nu(\mathbf{r}) \phi_\nu(\mathbf{r}+\mathbf{s}) d^2r \right) d^2s; \tag{B.15}$$

$$M_2 = \frac{2Q}{D^2} \sum_\nu \frac{\mu_\nu^{(2)}}{(k_\nu^2 + \alpha^2)^2}, \quad \mu_\nu^{(2)} = \int_{\mathbb{R}^2} s^2 \left( \int_\Omega \phi_\nu(\mathbf{r}) \phi_\nu(\mathbf{r}+\mathbf{s}) d^2r \right) d^2s. \tag{B.16}$$

Therefore, the effective correlation half-width in a finite domain is obtained as

$$\sigma_P^2 = \frac{1}{2} \frac{\sum_\nu \mu_\nu^{(2)} (k_\nu^2 + \alpha^2)^{-2}}{\sum_\nu \mu_\nu^{(0)} (k_\nu^2 + \alpha^2)^{-2}}. \tag{B.17}$$

This result corresponds precisely to equation (2.32) in the main text.

# Appendix C. Gaussianization of spatial correlations in multi-channel detection

In section 4.2, an equivalent Gaussian representation was introduced to quantify spatial sampling overlap in multi-channel measurements through a second-moment–preserving parametrization. Here we provide its explicit derivation in two dimensions, showing how it follows naturally from the analytic structure of the diffusion–relaxation covariance kernel and from the additivity of second moments under convolution.

We begin with the linear diffusion-relaxation operator $\mathcal{L} = -\nabla^2 + \alpha^2$, whose Green's function satisfies

$$\left(\nabla^2 - \alpha^2\right)\mathcal{G}_L(\mathbf{r}) = \delta(\mathbf{r}). \tag{C.1}$$

In two spatial dimensions, the solution is given by the modified Bessel function

$$\mathcal{G}_L(\mathbf{r}) = -\frac{1}{2\pi D} K_0(\alpha \mathbf{r}), \tag{C.2}$$

with Fourier transform

$$\tilde{\mathcal{G}}_L(\mathbf{k}) = -\frac{1}{D|\mathbf{k}|^2 + \Gamma} = -\frac{1}{D} \cdot \frac{1}{|\mathbf{k}|^2 + \alpha^2}. \tag{C.3}$$

Since overall amplitude plays no role in normalized moments, we introduce the normalized response kernel, defined in equation (B.4),

$$g(r) \equiv \frac{\mathcal{G}_L(r)}{\int \mathcal{G}_L} = \frac{\alpha^2}{2\pi D} K_L(\alpha r), \qquad \int g(r) d^2 r = 1. \tag{C.4}$$

interpreted as the spatial probability density associated with the diffusive response to a single, localized fluctuation.

Under the assumption of spatially $\delta$-correlated stochastic forcing, the equal-time covariance kernel takes the convolution form

$$\bar{C}_P(s) \propto (\mathcal{G}_L * \mathcal{G}_L)(s) \tag{C.5}$$

Factoring out the overall amplitude, we introduce the normalized covariance shape kernel, defined in equation (B.5),

$$h(s) = (g * g)(s) = \int_{\mathbb{R}^2} g(\mathbf{u}) g(\mathbf{u} + \mathbf{s}), \qquad \int h(s) d^2 s = 1. \tag{C.6}$$

Physically, it represents the spatial overlap of two statistically independent diffusive responses initiated at the same point. By construction, all information about the magnitude of fluctuations is absorbed into the prefactor, while the spatial structure of correlations is entirely encoded in $h(s)$.

In Fourier space, this relation simplifies to

$$\phi_h(\mathbf{k}) = \phi_g(\mathbf{k})^2, \qquad \phi_g(\mathbf{k}) = \frac{\alpha^2}{|\mathbf{k}|^2 + \alpha^2}. \tag{C.7}$$

To extract the spatial extent of correlations, we expand $\phi_g(\mathbf{k})$ at small wavevector,

$$\phi_g(\mathbf{k}) = 1 - \frac{|\mathbf{k}|^2}{\alpha^2} + O(k^4) \tag{C.8}$$

and invoke the general isotropic relation in $d$-dimensions,

$$\phi_f(\mathbf{k}) = 1 - \frac{|\mathbf{k}|^2}{2d}\langle r^2 \rangle + O(k^4). \tag{C.9}$$

For $d = 2$, we have

$$\langle r^2 \rangle_g = \frac{4}{\alpha^2} = 4\lambda_D^2. \tag{C.10}$$

Since $h = g * g$, the second moment adds,

$$\langle r^2 \rangle_h = 2\langle r^2 \rangle_g = 8\lambda_D^2. \tag{C.11}$$

This result shows that the spatial extent of the covariance kernel is set by the diffusive length scale $\lambda_D$, independently of any assumed Gaussian form.

To obtain a compact parametrization, we introduce an equivalent Gaussian kernel $G(r; \sigma_h) \approx h(r)$ whose second moment matches that of $h$,

$$2\sigma_h^2 = \langle r^2 \rangle_h \implies \sigma_h^2 = \frac{4}{\alpha^2} = 4\lambda_D^2 \tag{C.12}$$

This Gaussian does not approximate the detailed shape of $h$, but preserves its second moment by construction.

In multi-channel measurements, spatial correlations arise from two statistically independent contributions: the intrinsic fluctuation covariance encoded in $h$, and the finite spatial extent of the measurement weights. For a Gaussian measurement weight, as is appropriate for typical laser probe beams with diffraction-limited or weakly aberrated intensity profiles,

$$W(r) = \exp\left(-\frac{r^2}{2w^2}\right) \quad \text{(C.13)}$$

the covariance involves the squared weight $R(r) = W(r)^2$, reflecting the bilinear appearance of the measurement functional in the covariance. The second moment evaluates to

$$\langle r^2 \rangle_R = w^2 \quad \text{(C.14)}$$

which characterizes the effective spatial extent of the measurement contribution to the covariance.

Because convolution corresponds to the addition of variances, the total effective kernel has second moment

$$\langle r^2 \rangle_{\text{tot}} = \langle r^2 \rangle_R + \langle r^2 \rangle_h = w^2 + 8\lambda_D^2. \quad \text{(C.15)}$$

Defining an equivalent Gaussian representation $G(r; \sigma_{\text{eff}})$ via $2\sigma_{\text{eff}}^2 = \langle r^2 \rangle_{\text{tot}}$, we obtain the effective sampling half-width

$$\sigma_{\text{eff}}^2 = \frac{w^2}{2} + 4\lambda_D^2, \quad \text{(C.16)}$$

which is equation (4.3) of the main text.

Spatial sampling overlap is thus determined by the second central moment of the effective covariance kernel. The Gaussian representation serves only as a moment-preserving parametrization, and all resulting scaling relations are governed by the low-order covariance spectrum.